\documentclass[aps,twocolumn,superscriptaddress,letterpaper]{revtex4-1}
\usepackage{graphicx,amssymb,bm,natbib,amsfonts,amsmath}
\usepackage{hyperref}

\begin{document}

\title{Crossover region between nodal and anti-nodal states at the Fermi level of optimally and overdoped Bi$_{2}$Sr$_{1.6}$Nd$_{0.4}$CuO$_{6+\delta}$}

\author{D.R. Garcia}
\affiliation{Department of Physics, University of California,
Berkeley, CA 94720, USA}
\affiliation{Materials Sciences Division,
Lawrence Berkeley National Laboratory, Berkeley, CA 94720, USA}
\author{J. Graf}
\affiliation{Materials Sciences Division,
Lawrence Berkeley National Laboratory, Berkeley, CA 94720, USA}
\author{C. Jozwiak}
\affiliation{Materials Sciences Division,
Lawrence Berkeley National Laboratory, Berkeley, CA 94720, USA}
\author{C.G. Hwang}
\affiliation{Materials Sciences Division,
Lawrence Berkeley National Laboratory, Berkeley, CA 94720, USA}
\author{H. Eisaki}
\affiliation{AIST Tsukuba Central 2, Umezono, Tsukuba, Ibaraki, 305-8568, Japan}
\author{A. Lanzara $^\dagger$}
\affiliation{Department of Physics, University of California,
Berkeley, CA 94720, USA}
\affiliation{Materials Sciences Division, Lawrence Berkeley National Laboratory, Berkeley, CA 94720, USA}

\date{\today}

\begin{abstract}

We have studied Bi$_{2}$Sr$_{1.6}$Nd$_{0.4}$CuO$_{6+\delta}$ using Angle Resolved Photoemission Spectroscopy in the optimal and overdoped regions of the phase diagram.  We identify a narrow crossover region in the electronic structure between the nodal and antinodal regions associated with the deviation from a pure d-wave gap function, an abrupt increase of the quasiparticle lifetime, the formation of Fermi arcs above T$_c$, and a sudden shift of the bosonic mode energy from higher energy, $\sim${60meV}, near the nodal direction, to lower energy, $\sim${20meV}, near the antinodal direction.  Our work underscores the importance of a unique crossover region in the momentum space near E$_F$ for the single layered cuprates, between the nodal and antinodal points, that is independent of the antiferromagnetic zone boundary.

\end{abstract}

\maketitle


Understanding the near-E$_F$ electronic band structure is essential to making sense of the superconducting cuprate phase diagram. Increasingly, studies on these systems are suggesting that the Fermi surface (FS) may be better thought of as divided into regions along k$_F$\cite{NormanPerspectives}\cite{Perali}\cite{Bianconi}. For example, the partial gapping of the FS in the pseudogap (PG) phase results in the unusual formation of ungapped regions which appear as disconnected arcs or ``Fermi Arcs" (FA).  Although these FA appear to scale with T*\cite{KanigelNature}, the initial region in which they form near the nodal point (where $\Delta$=0 for T$<$T$_c$) above T$_c$ remains mysterious.  Additionally, recent experiments have pointed out that the superconducting (SC) gap departs from a simple d-wave form, and opposing gap trends exist in different areas of the FS. This points to the existence of an additional energy scale within the SC gap function\cite{Harris}\cite{Deutscher}\cite{He}\cite{KondoNature}\cite{Hashimoto}\cite{Kondo}, though this position still remains a subject of debate\cite{HBYang}\cite{KanigelPRL08}\cite{JMeng}.  If true, it also would suggest a division of the FS where electronic states near the node are uniquely related to SC while the antinode (AN) (where $\Delta$=Max for T$<$T$_c$) would be related to an additional, potentially competing, phenomenon corresponding to the PG phase\cite{KondoNature}\cite{Emery}\cite{Castellani}\cite{NLSaini}\cite{Tanaka}\cite{W.S.Lee}.  STM work has suggested another potential crossover region on the Fermi surface associated with the antiferromagnetic zone boundary, characterized by the disappearance of the coherent quasiparticle (QP) peak\cite{Davis}.  Also, bosonic modes which introduce ``kinks'' in the near-E$_F$ band structure may define a key region of the FS.  Recent work finds a shift in the kink energy towards low binding energy correlated with the softening of the Cu-O bond stretching (BS) phonon at a critical FS nesting wavevector\cite{Jeffpaper}.  Finally, recent quantum oscillation data\cite{Doiron-Leyraud}\cite{Sebastian} have suggested that the FS is indeed separated into hole and electron pockets providing another crossover region for the low energy QP.  Although controversial, evidence for an hole pocket on the FS has been recently provided by photoemission experiments\cite{NewJMeng}.  Thus, it is increasingly important that we continue to explore the physics of these different regions.

\begin{figure}
\includegraphics[width=8.25 cm]{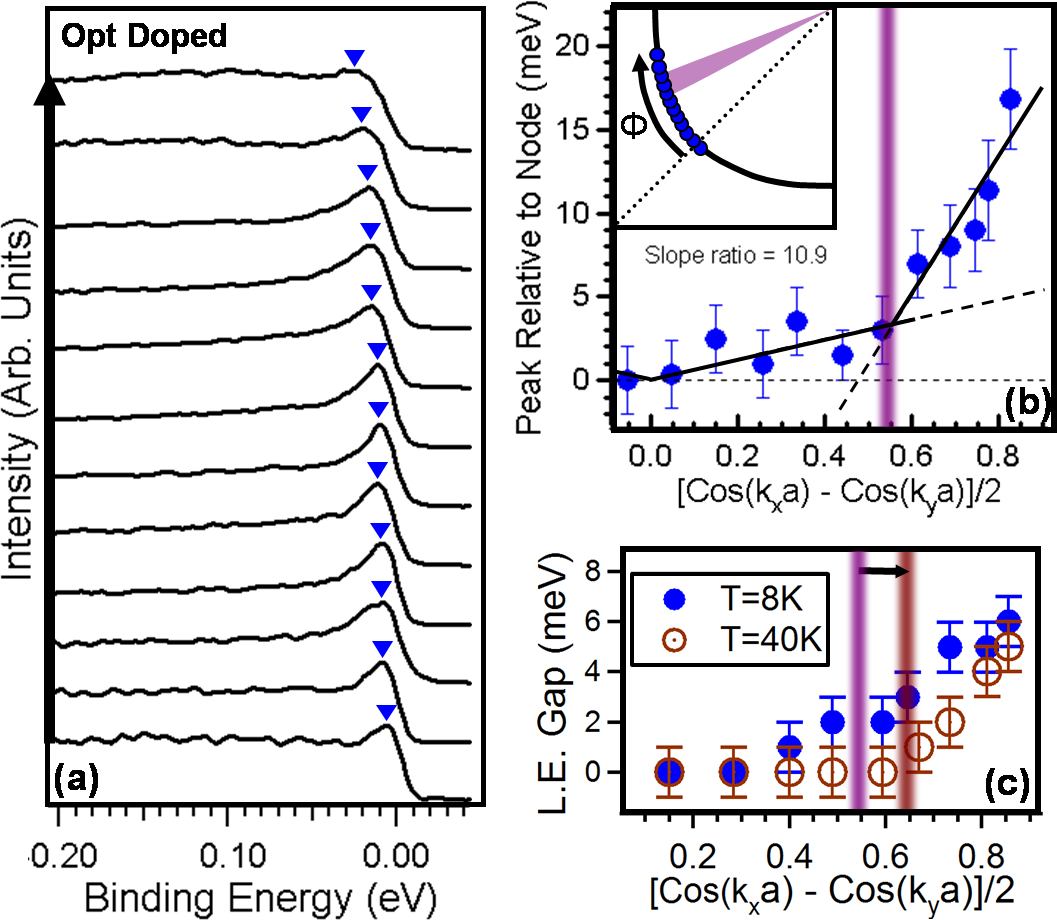} 
\caption{(color online) (a) EDC spectra taken at T=8K on the OP Nd-Bi2201 at k$_F$ along the band structure. (b) The QP peak binding energy positions indicated by the blue triangles in (a) and shifted relative to the nodal point peak.   Inset shows locations of these EDC spectra at E$_F$.  The lines are guides to the eye, indicating a deviation from a pure d-wave gap function when plotted on this abscissa.  The violet shaded area in all these panels indicates a point 18$^{\circ}\pm $1$^{\circ}$ away from the nodal point. (c) Leading edge gap data for both SC and PG phases.  Here the violet region is shifted to include the expected FA scaling for T=40K (indicated by the brown line).}   
\end{figure}	
    
    In this paper, we do high resolution Angle Resolved Photoemission Spectroscopy (ARPES) measurements on single crystal samples of Bi$_{2}$Sr$_{1.6}$Nd$_{0.4}$CuO$_{6+\delta}$ (Nd-Bi2201) in both optimally doped (OP) (T$_c$=27.5K) and overdoped (OD) (T$_c$=10K) regions of the hole-doped phase diagram.  ARPES, being a direct probe of the electronic band structure, is uniquely suited to these studies and has been instrumental in most of the aforementioned works.  Being a single layered compound, Nd-Bi2201 lacks the bilayer splitting of its well studied cousin, Bi2212, allowing for easier interpretation of the FS band structure.  The presence of the Nd$^{3+}$ lanthanide introduces strain due to lattice mismatch and decreases the T$_c$ from the unsubstituted parent compound\cite{Eisaki}.  Measurement of T* places it well above the peak T$_c$ for the OP sample\cite{Lavrov} and suggest the PG phase is stronger in Nd-Bi2201 than nonstrained Bi2201 systems\cite{Hashimoto}.                    
                  
	  	We find evidence for a narrow crossover region in the band structure of Nd-Bi2201 separating the nodal from the AN states.  This point in momentum space appears to be correlated with: a) Discontinuity in the d-wave character of the gap, both for optimal and overdoping; b) Abrupt increase of the QP lifetime, $\Gamma$; c) A red shift of the bosonic kink energies, consistent with the shift observed in the unstrained La-Bi2201\cite{Jeffpaper}.  The location of this crossover region appears consistent with the ends of the FA as it forms just above T$_c$ in the PG phase of the OP sample. These results suggest that whatever the mechanism is that drives the formation of this crossover region, it marks an abrupt change in many physical quantities, suggesting a close connection with the SC state.

    Samples of single crystal Nd-Bi2201 were grown using the traveling solvent floating-zone technique\cite{Eisaki}.  Synchrotron ARPES data were taken at Beamline 5.4 at the Stanford Synchrotron Radiation Laboratory, using a Scienta R4000 analyzer.  A total energy resolution of $<$13meV was achieved for data taken on OP (T$_{c}$=27.5K) Nd-Bi2201 while a resolution of $<$8meV was achieved for data taken on OD (T$_{c}$=10K) Nd-Bi2201.  In both cases, the angular resolution was better than 0.35$^{\circ}$.  Fresh sample surfaces were prepared by cleaving the sample {\em in situ} at a base pressure $<$5x10$^{-11}$ torr at low temperatures.

  Figure 1 shows data taken on the OP Nd-Bi2201 sample.  Panel a provides a stack of Energy Distribution Curves (EDC) taken at k$_F$ illustrating the evolution of the QP peak along the band structure manifold in momentum space.  The binding energies of the QP peak can quantify the gap function so long as the peak can be clearly observed over the rising background. In light of the functional form of a d$_{x^2-y^2}$ wave gap $\Delta_{k}=\Delta(cos(k_{x}a)-cos(k_{y}a))/2$, we can plot these energies with respect to this function as commonly done\cite{He}\cite{Hashimoto}\cite{W.S.Lee}, revealing potential deviations from a single d-wave gap.  The result is panel b where a clear deviation from a pure d-wave is observed, similar to other cuprate studies\cite{Harris}\cite{He}\cite{KondoNature}\cite{Hashimoto}\cite{Kondo}\cite{W.S.Lee}. The proposed location of the crossover is indicated in violet.

To explore the connection between this crossover and FA formation, Fig.~1c shows the gap along k$_F$ above T$_c$.  After compensating for the expected T/T* FA temperature dependence\cite{KanigelNature} (violet to brown line in the figure) and using T*=140K\cite{Hashimoto}\cite{Lavrov}, we can compare the location of the crossover to the size of the FA when it first forms above T$_c$.  We find that the location of the crossover lies approximately where this initially formed FA terminates (the gap finally opens), a correlation which seems consistent with other published Bi2212 ARPES studies\cite{W.S.Lee}.  This suggests a potential connection between FA formation in the PG phase and the apparent two d-wave gaps observed in the SC phase.

\begin{figure}
\includegraphics[width=8.15 cm]{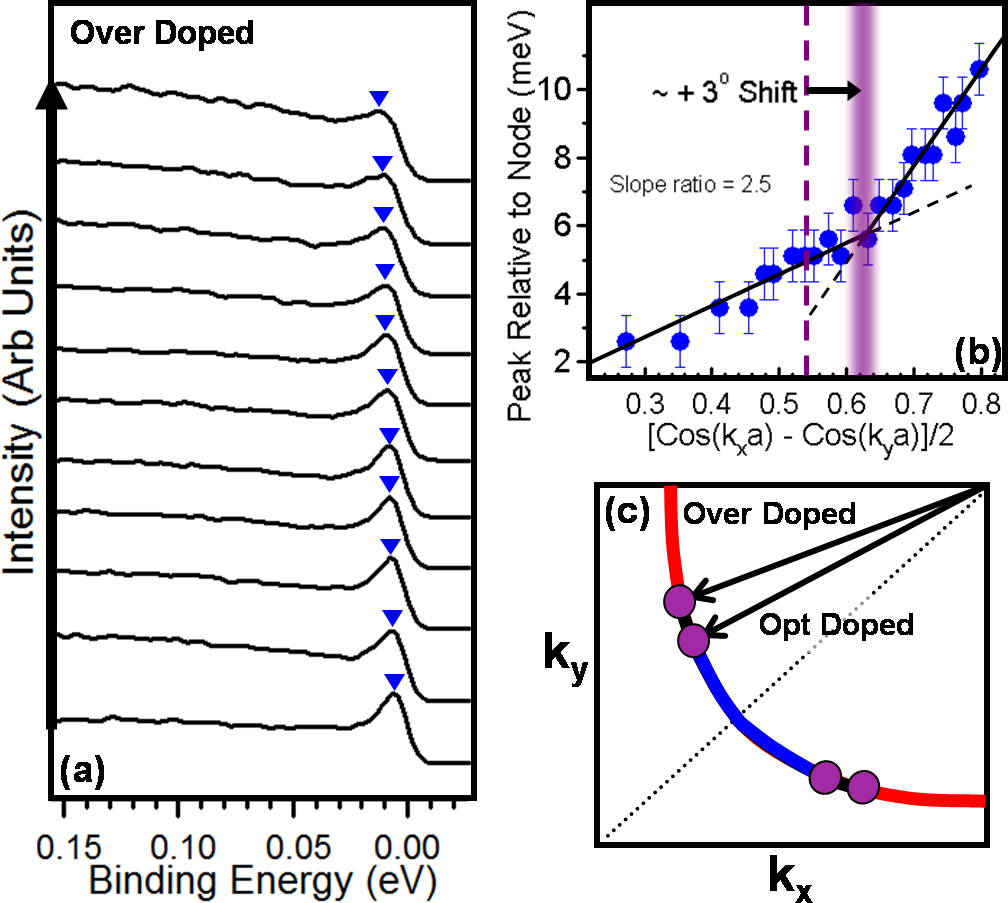}
\caption{(color online) (a) Sampling of EDC spectra taken at T=8K on the OD Nd-Bi2201 along k$_F$. (b) QP peak binding energy positions indicated by the blue triangles in (a), shifted relative to the nodal point peak, for all EDCs measured along E$_F$.  The lines are guides to the eye indicating a deviation from a pure d-wave gap function.  The violet shaded area in these panels indicates a point shifted by $\sim$3$^{\circ}$ from Fig.~1 (dashed line). (c) Cartoon illustrating the regions (blue and red) identified in Figures 1 and 2 and small shift in location of the crossover (violet circles) between the dopings.}  
\end{figure}

Figure 2 turns our attention to the OD Nd-Bi2201 sample.  Panel a shows the evolution of the QP peak at k$_F$ for a sampling of EDCs along the FS as done in Figure 1.  Plotting the peak binding energies for all EDCs in the manner of Fig.~1b, there remains evidence of deviation from a pure d-wave gap function although less pronounced with the ratio of the two slopes being $\sim$2.5 for the OD compared to $\sim$10.9 for the OP sample.  Since observed deviations from a pure d-wave gap tend to disappear with overdoping in the cuprates, its continued presence in our data may be due to the increased strain within the lattice.  Above T$_c$, we were unable to find clear evidence of Fermi arcs in our data for comparison, which is consistent with the OD region of the phase diagram where T* becomes closer to T$_c$ or below it.

\begin{figure}
\includegraphics[width=8.7500 cm]{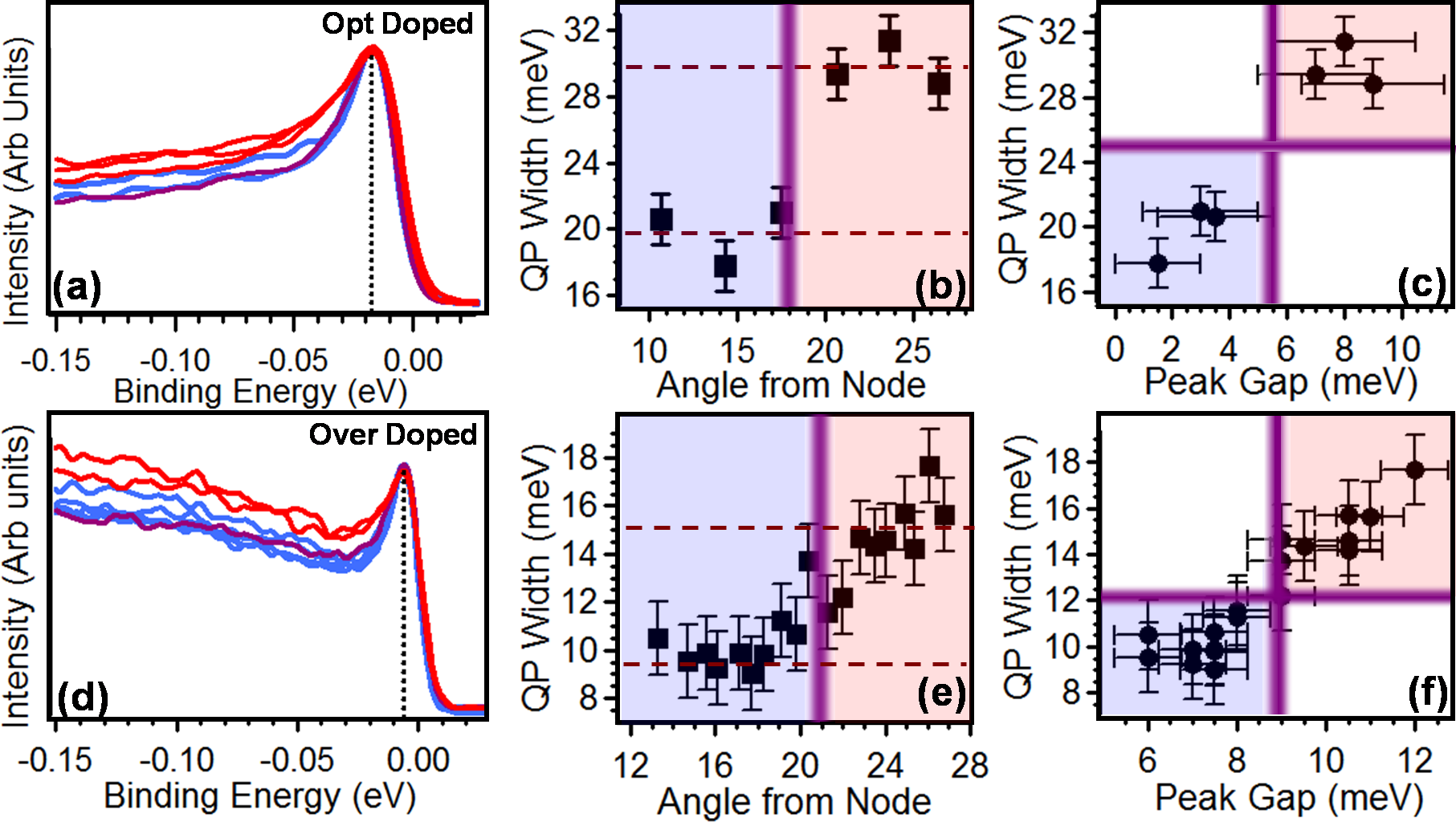}
\caption{(color online) QP peak width data taken in the SC phase for (a)-(c) OP and (d-f) OD samples. (a,d) Selected EDCs from Fig.~1a and 2a, respectively, with their peaks aligned, normalized to QP peak height, taken on both sides of the crossover region.  Blue (red) curves indicate the nodal (AN) side of the crossover.  The violet curve sits right on the crossover region.  (b,e) Fitted QP widths quantifying the energy broadening through the crossover region with error determined from fit.  (c,f) QP widths plotted versus their associated peak gap from Figures 1 and 2.}
\end{figure}

		Fig.~2c provides a simple cartoon to illustrate the emerging picture. States near the node (indicated in blue) are separated by a crossover region (violet circles) from states closer to the AN point (indicated in red.)  Our data suggests this crossover region may be shifting towards the AN with overdoping. We can estimate the crossover location at 18$^{\circ}\pm $1$^{\circ}$ from the nodal point or near ($\pm$$\pi$/4.3a$_0$, $\pm$$\pi$/1.6a$_0$) $\pm$ 5$\%$ in momentum space for the OP sample.  In the OD compound, the crossover occurs at 21$^{\circ}\pm $1$^{\circ}$ or near ($\pm$$\pi$/5a$_0$, $\pm$$\pi$/1.6a$_0$) $\pm$ 5$\%$, shifted ($\sim$3$^{\circ}$) towards the AN, a trend consistent with other work\cite{W.S.Lee}.  At this point, two issues should be clarified.  First, it is not at all clear that this crossover necessarily represents a finite region of states separate from the nodal and AN states.   Second, although the cartoon includes all the states beyond the crossover as red, we are unable to fully explore all of the states nearest the AN point.  This means there could conceivably be an additional crossover which happens much nearer the AN point.  In which case this red region may in fact be a finite middle region and not extend all the way to the Brillouin zone edge\cite{Bianconi}.

With this picture, we look for further signatures of the crossover affecting the electronic states in this region of momentum space.  Returning to the individual EDCs, we find evidence of sudden changes in the QP lifetime occurring near the crossover region, compared to the gradual increase in lifetime observed in the normal phase between the nodal and antinodal points\cite{Kaminski}.  Figure 3 summarizes these results for both dopings.  In panel a, peak-aligned EDCs from near the crossover region in the OP sample are stacked with blue and red curves corresponding to the angular region indicated in the Fig.~2c cartoon. The violet curve is associated with a data point effectively at the crossover.  This suggests that the electronic states closer to the node (blue) are suddenly altered upon passing through the violet crossover region towards AN states (red) along the near E$_F$ band structure.  One concern that may arise could be discerning between a widening peak and simply a weakening peak over the inelastic background.  First, even if this were the case, the sudden drop in QP spectral weight could then be interpreted as the indicator of passing through the crossover region, still affirming the crossover's affect on the spectral lineshape.  Secondly, we can better quantify this change by extracting the QP lifetime-related energy width, $\Gamma$, through fitting the spectra to a resolution broadened spectral function using the well-known self-energy\cite{OldNorman} $\Sigma(k_{F},\omega)=-i\Gamma +\Delta^{2}/\omega$ within the Green's function, and sitting on a variable inelastic background.  The results are displayed in panel b.  Again, we find the change in QP width occurs near the violet crossover region for the OP sample as identified in Figure 1.

\begin{figure*}
\includegraphics[width=18 cm]{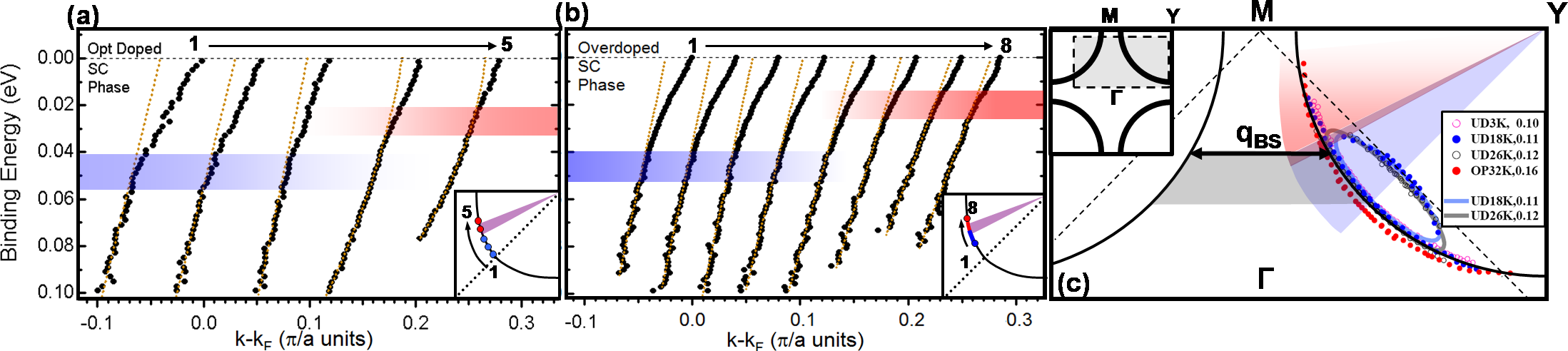}
\caption{(color online) (a-b) Fitted band dispersions near E$_F$, horizontally shifted, from ARPES data on (a) OP and (b) OD samples in the SC phase.  Data is taken near the crossover region, indicated in the respective insets, with dashed lines as guides to the eye, and high and low energy kinks indicated by blue and red, respectively.  c) FS cartoon summarizing the different issues associated with the crossover region.  Colored angular regions correspond to the OP sample's division of the FS as suggested by Fig.~2c.  Grayed area corresponding to the region of the softened Cu-O BS phonon mode from Ref.~\cite{Jeffpaper}. Circular data points come from Ref.~\cite{NewJMeng} indicating the FS pocket in La-Bi2201.}   
\end{figure*}

    We can examine this effect in the OD samples in the same manner.  In panel d, the QP peak width appears sharper, consistent with an increased doping in the bismuth cuprates\cite{Harris}\cite{Yoshida}. We can make out a change in the spectra associated with passing through the violet crossover EDC into the red EDCs.  As before, we can use fitting techniques to better explore the QP lifetime as plotted in panel e with the violet crossover region identified in Figure 2.  This finer survey of $\Gamma$ again suggests that the QP width begins to rise as we pass through the crossover although slightly more gradual than we observed for the OP sample.  Panels c and f plot the gap data from Figs.~ 1b and 2b with the QP width data to underscore the correlation between the identified regions.  This decrease in QP lifetime as electronic states approach the AN point seems to be in agreement with studies in the literature suggesting the gap near the AN is due to some competing phase\cite{Emery}\cite{Castellani}\cite{NLSaini}\cite{KondoNature}\cite{Tanaka}\cite{W.S.Lee}.  Indeed, we find that the location of our crossover region for the OP sample and the general doping trend corresponds well with the maximum in the coherent peak weight, W$_{CP}$, seen by Kondo et al.~on the unstrained Bi2201\cite{KondoNature} where a competitive relationship between the SC and PG phases is suggested.

A final avenue of study is to consider how the bosonic renormalizations or kinks may be affected along k$_F$.  This is particularly significant because this bosonic shift may be connected to the softening of a particular phonon mode with a characteristic wavevector seen by INS in La-Bi2201\cite{Jeffpaper}.  Indeed, this shifting of the kink energy from the node to the AN has long been observed in the double layered bismuth cuprates\cite{Cuk}\cite{Gromko}\cite{Terashima}. In Fig.~4a, we show the fitted dispersions taken on the OP Nd-Bi2201 sample in the SC phase.  In continuity with La-Bi2201, we find evidence of a shift in the kink as indicated by the blue and red shaded regions, although the energy of both modes is slightly less.  We associate the blue higher energy kink with a similar feature seen throughout the cuprates\cite{Alessandra}.  This shift is generally consistent with the location where we expect the INS wavevector from La-Bi2201 to nest the OP Nd-Bi2201 FS. Additionally interesting is comparing the location of the kink shift to the crossover region.  We find the transition between the kink energies is roughly between 14$^\circ$ and 18$^\circ$ off the nodal point, close enough to the crossover region to suggest a connection.

In panel b, we show the fitted dispersion for the OD sample. The same shift in the bosonic kink is observed between the two energy scales and its location appears roughly between 18$^\circ$ and 20$^\circ$ off the nodal point, again placing it close to the crossover as observed in the OD sample. One issue that does arise is that nesting the OD FS with the INS wavevector from the La-Bi2201 work does not correspond well to the location of this shift.  However, with no sufficient quality INS data on OD Bi2201 nor on the strained Nd-Bi2201, there is no reason to suppose the associated wavevector of the phonon softening would remain the same magnitude in this system.

Fig.~4c provides a cartoon to bring together the several threads associated with this crossover region and its potential origin.  All of the effects discussed are sufficiently far from the antiferromagnetic zone boundary (dashed line) to preclude such an explanation for what we observe.  The proximity of the bosonic shift to the crossover region seems to suggest a connection between the crossover and the softened Cu-O bond stretching phonon (region shaded gray).  Indeed, its associated softening wavevector, $\bf{q_{BS}}$, could connect the FS near the location of our crossover region. This region also falls close to other Bi2201 work on Fermi arc formation\cite{Nakayama} and the aforementioned evolution in coherent peak spectral weight\cite{KondoNature}. Additionally, this region of k-space has connections to recent work of J.~Meng et al.~\cite{NewJMeng} where data from La-Bi2201 suggest the nodal point forms a hole pocket.  Data from their work mapping the FS is included in panel c.  The separation of the cupate FS into hole and electron pockets had been strongly proposed by quantum oscillation\cite{Doiron-Leyraud}\cite{Sebastian}.  We find it significant that the colored angular regions place the violet OP crossover region very close to the apparent closing of the proposed hole pocket. Still, the correlation between the ARPES work and the pockets observed by quantum oscillation remains controversial.

	   In conclusion, we have taken ARPES data on Nd-Bi2201 at both OP and OD regions of the phase diagram.  We have found evidence of a narrow crossover region associated with: 1) A transition between two d-wave like gaps with different energy scales, 2) The initial formation of the FA in the PG phase seen in our OP data, 3) An anomalous increase in the QP lifetime energy, and 4) The shift in the binding energy of the kink in the near E$_F$ band structure. The transition between these two kink energy scales suggests a potential connection between the softening Cu-O bond stretching mode and the crossover. 
	   	   
$^\dagger$ Electronic address: alanzara@lbl.gov

We would like to thank S. Wilson, C. Rotundu, D.-H. Lee, C. Smallwood, and D. Siegel for helpful discussions.  
This work was supported by the Director, Office of Science, Office of Basic Energy Sciences, Materials Sciences and Engineering Division, of the U.S. Dept.~of Energy under Contract No. DE-AC02-05CH11231. Portions of this research were carried out at the Stanford Synchrotron Radiation Lightsource, a national user facility operated by Stanford University on behalf of the U.S. Department of Energy, Office of Basic Energy Sciences.

\end{document}